\documentclass[aps,pra,reprint,twocolumn,amsmath,amssymb,floatfix]{revtex4-2}

\usepackage{bm}
\usepackage{graphicx}
\usepackage{dcolumn}
\usepackage[export]{adjustbox}

\begin{document}

\title{Critical fields of superconductors with magnetic impurities}

\author{V. G. Kogan}
\affiliation{Ames Laboratory - US DOE and Department of Physics \& Astronomy, Iowa State University, Ames, IA 50011, U.S.A.}

\author{R. Prozorov}
\affiliation{Ames Laboratory - US DOE and Department of Physics \& Astronomy, Iowa State University, Ames, IA 50011, U.S.A.}

\begin{abstract}
The upper critical field $H_{c2} $, the field $H_{c3}$ for nucleation of the surface superconductivity, and the thermodynamic $H_c $ are evaluated within the weak-coupling theory for the isotropic s-wave case and arbitrary transport and pair-breaking scattering. We find that for the standard geometry of a half-space sample in a magnetic field parallel to the surface, the ratio ${\cal R}=H_{c3}/H_{c2}$ is within the window  $1.55\lesssim {\cal R}\lesssim 2.34$, regardless of temperature, magnetic or non-magnetic scattering. While the non-magnetic impurities tend to flatten the ${\cal R}\left(T\right)$ variation, the magnetic scattering merely shifts the maximum of ${\cal R}\left(T\right)$ to lower temperatures.   Surprisingly, while reducing the  transition temperature,  magnetic scattering has
a milder impact on ${\cal R}$  than the non-magnetic scattering. The surface superconductivity is quite robust; in fact, the ratio ${\cal R}\approx 1.7$ even in the gapless state. We used Eilenberger's energy functional  to evaluate the condensation energy $F_c$ and the thermodynamic critical field $H_c$ for any temperature and scattering parameters.    
By comparing $H_{c2} $ and $H_{c}$, we find that unlike the transport scattering, the pair-breaking  pushes materials  toward type-I behavior. We find a peculiar behavior of $F_c$ as a function of the pair-breaking scattering parameter at the low-$T$ transition from gapped to gapless phases, which has recently been associated with the topological transition in the superconducting density of states.
\end{abstract}

\date{Submitted: 23 June 2022}

\maketitle

\section{Introduction}

The question of the critical fields $H_{c2}$ and $H_{c3}$ is practically relevant since it is directly related to the critical temperature $T_c(H)$ where the superconductivity emerges in the applied magnetic field $H$. In type-II materials at a fixed $T$, in the increasing field, the  vortex phase in the bulk is destroyed at $H_{c2}$, but the superconductivity may survive in a coherence-length-deep surface layer up to $H_{c3}$. 

In recent decades the interest in limiting fields was further fueled by the significant progress in superconducting resonator cavities used in particle accelerators \cite{Gurevich2012,Liarte2017}, and even more recently, in the hardware for superconducting circuits--based quantum computing \cite{Devoret2013,Rigetti2012}. Of particular interest are effects of a disorder that influences cavities quality factors and superconducting qubits coherence times \cite{Gurevich2012,Sauls2019}.

The ratio ${\cal R}=H_{c3}/H_{c2}$ for the applied field parallel to the surface of isotropic superconducting half-space has been evaluated by Saint-James and DeGennes (SJDG) \cite{DG} by solving the linearized Ginzburg-Landau equations for the order parameter $\Delta$ subject to the boundary condition of vanishing normal gradient, $\nabla_n\Delta=0$,  at the sample surface. Their  seminal result is ${\cal R}=1.695$.
Since then, the surface superconductivity has been observed in many experiments, but the ratio $\cal R$ varied depending on surface quality, sample anisotropy, set up geometry, scattering, and temperature \cite{SJ,DGbook,Tomasch,hemp,Strong}. Theoretically, effects of material anisotropy have been discussed in \cite{Max}, where it was shown that for sufficiently high anisotropy for some surface orientation, the ratio ${\cal R}$ may fall under unity. In other words, surface superconductivity does not exist. 

An interesting development came recently, showing that within microscopic BCS theory, ${\cal R}(T)$ has a maximum at intermediate temperatures, which, however, disappears with increasing transport scattering \cite{Levch}.
In this contribution, we extend this study   to the case when both magnetic
and non-magnetic scattering channels are present. The
discussion is limited to isotropic material with isotropic Fermi surface
and s-wave order parameter. Given that  $H_{c2}$ is enhanced by non-magnetic  transport scattering whereas suppressed by magnetic impurities  \citep{KP-2013}, the question of the effect of magnetic impurities on $H_{c3}$ is not obvious.

 \section{The problem of $\bm{H_{c2}}$ and $\bm{H_{c3}}$}
 
Consider an isotropic material with both magnetic and
non-magnetic scatterers; $\tau_m$ and $\tau$ are the
corresponding average scattering times. The problem of the second-order phase transition at 
$H_{c2}$  and $H_{c3}$ is addressed on the basis of Eilenberger's quasiclassical
version of Gor'kov's equations for normal and anomalous Green's
functions $g$ and $f$. At the 2nd order phase transition, $g=1$ and we are left with a linear equation for $f$ \cite{Eil,KPMishra}:
\begin{eqnarray}
(2\omega^+ +{\bf v}\cdot {\bm \Pi})\,f=2\Delta
/\hbar+ \langle f\rangle/\tau^- \,,\label{E1}\\
\omega^+=\omega + \frac{1}{2\tau^+}\,,\qquad\frac{1}{\tau^\pm}=\frac{1}{\tau
}\pm\frac{1}{\tau_m}\,.
\label{om+}
\end{eqnarray}
Here, ${\bm v}$ is the Fermi velocity, ${\bm  \Pi} =\nabla +2\pi i{\bm 
A}/\phi_0$ with    the vector potential $\bm A$ and  the flux quantum $\phi_0$. $\Delta ({\bm  r})$ is  the order parameter; 
 Matsubara frequencies are defined by $ \omega=\pi T(2n+1)$
with an integer $n$; in the following (except  some final results) we set $\hbar=1$; $\langle...\rangle$ stand for  averages over
the Fermi surface.  
Solutions $f$ of  Eq.\,(\ref{E1}) along with   $\Delta$ should satisfy the  self-consistency equation:
\begin{equation}
\frac{\Delta}{2\pi  T}\ln\frac{T_{c0}}{T}=\sum_{\omega>0}\left(\frac{\Delta}
{ \omega }-\langle f\rangle\right)\,,
\label{selfcons0}
\end{equation}
where $T_{c0}$ is the critical temperature in the absence of pair-breaking scattering. 

Helfand and Werthamer \cite{HW} had shown that at the 2nd order phase transition at $H_{c2}$, the order parameter satisfies a linear equation
\begin{eqnarray}
{\bm \Pi}^2\Delta =k^2\Delta\,. 
\label{HW}
\end{eqnarray}
It was realized later that this equation holds at any 2$^{nd}$ order transition from  normal to superconducting state away of $H_{c2}$, e.g. in proximity systems or at $H_{c3}$,  provided $k^2=-1/\xi^2$ satisfies the self-consistency equation of the theory \cite{85,KN}. It turned out that the coherence length so evaluated depends not only on temperature and scattering but also on the magnetic field (except in the dirty limit or near $T_c$). In this sense, Eq.\,(\ref{HW}) in fact differs from the {\it linearized} Ginzburg-Landau equation that forms the basis for   SJDG  prediction of the surface superconductivity at $H_{c3}$ \cite{DG}. It is worth noting that if $\xi$ would have been $H$ independent, the ratio $H_{c3}/ H_{c2}$  would have been constant equal 1.695  at all temperatures. As was shown in \cite{Levch},  this is not so (except for the dirty limit). 

Thus, the order parameters at both $H_{c2}(T)$ and $H_{c3}(T)$ satisfy the same Eq.\,(\ref{HW}). The difference, however, comes from  boundary conditions: $\Delta(\bm r)$ should be finite everywhere for $H_{c2}$, whereas   $\nabla_n\Delta(\bm r)=0$ at the sample surface for $H_{c3}$ ($\nabla_n$ is the  gradient of $\Delta$ along the normal to the sample surface). 

At any 2$^{nd}$ order phase transition, $\Delta\to 0$ and one can deal with linear Eq.\,(\ref{E1}). Repeating the derivation  of Ref.\,\cite{85}, one finds (see the outline in Appendix A):
\begin{equation}
\langle f\rangle = \Delta\, \frac{2\tau^- S}{2\omega^+\tau^- -S} \,,
\label{F}
\end{equation}
where $S$ is given by a series
\begin{eqnarray}
S &=&\sum_{j,m=0}^\infty \frac{(-q^2)^j}{j!(2m+2j+1)}
\left(\frac{(m+j)! }{m!}\right)^2
\left(\frac{\ell^+ }{\beta^+}\right)^{2m+2j} 
\nonumber\\
&\times&  \prod_{i=1}^{m}\left[k^2+(2i-1)q^2\right],\qquad q^2=\frac{2\pi H}{\phi_0}\,,
\label{Series}
\end{eqnarray}
where
\begin{eqnarray}
  \ell^+=v\tau^+,\qquad  \beta^+=1+2\omega \tau^+\,.
\label{ell+}
\end{eqnarray}

This sum can be transformed to an integral, which is more amenable for the numerical work \cite{KN}:
\begin{eqnarray}
S&=&  \sqrt{\frac{\pi}{u}} \int_0^1\frac{d\eta\,(1+\eta ^2)^\sigma}{(1-\eta ^2)^{\sigma+1}}
 \left[{\rm erfc}\frac{\eta }{\sqrt{u}}-\cos(\pi\sigma) {\rm erfc}\frac{1}{\eta \sqrt{u}}\right]. \nonumber\\
\label{SKN}
\end{eqnarray}
Here,
\begin{eqnarray}
u= \left( \frac{q \ell^+}{\beta^+}\right)^2 =\frac{h}{[P^++t(2n+1 )]^2}  \,,    
\label{uKN}
\end{eqnarray}
where the reduced field $h$, temperature $t$, and the scattering parameter $P^\pm$ are introduced:
\begin{eqnarray}
h= H \frac{\hbar^2v^2}{2\pi\phi_0T_{c0}^2}\,,\quad t=\frac{T}{T_{c0}},\quad P^\pm=\frac{\hbar}{2\pi T_{c0}\tau^\pm} \qquad
\label{h,P}
\end{eqnarray}
($\hbar$ is written explicitly to stress that $h$ and $P^\pm$ are dimensionless).

The parameter $\sigma$ is defined as
\begin{eqnarray}
    \sigma=\frac{1}{2}\left(\frac{k^2}{q^2}-1\right)= -\frac{1}{2}\left(\frac{1}{q^2\xi^2}+1\right)\,.
\label{sig}
\end{eqnarray}
At $H_{c2}$, $\sigma=-1$ and  
\begin{eqnarray}
S (u)=  \sqrt{\frac{\pi}{u}} \int_0^1\frac{d\eta }{ 1+\eta^2}
 \left[{\rm erfc}\frac{\eta}{\sqrt{u}}+ {\rm erfc}\frac{1}{\eta \sqrt{u}}\right]. \qquad 
\label{Sc2}
\end{eqnarray}
  
Near $T_c$, the order parameter satisfies linearized GL equation $-\xi^2\Pi^2\Delta=\Delta$, and at $H_{c3}$ SJDG obtained $\xi^2q^2=1.695$ \cite{DG}. 

Therefore, at $H_{c3}$ we get
\begin{eqnarray}
 \sigma = -\frac{1}{2}\left(\frac{1}{1.695}+1\right)\approx  -0.795\,. 
\label{sigma}
\end{eqnarray}
Thus, to calculate $H_{c3}(t,P,P_m)$ with the help of Eqs\,(\ref{selfcons0}) and  (\ref{F}) one has to use   $S$ of Eq.\,(\ref{SKN}) with $ \sigma = -0.795 $. 

\begin{figure}[tb]
\includegraphics[width=8cm,left] {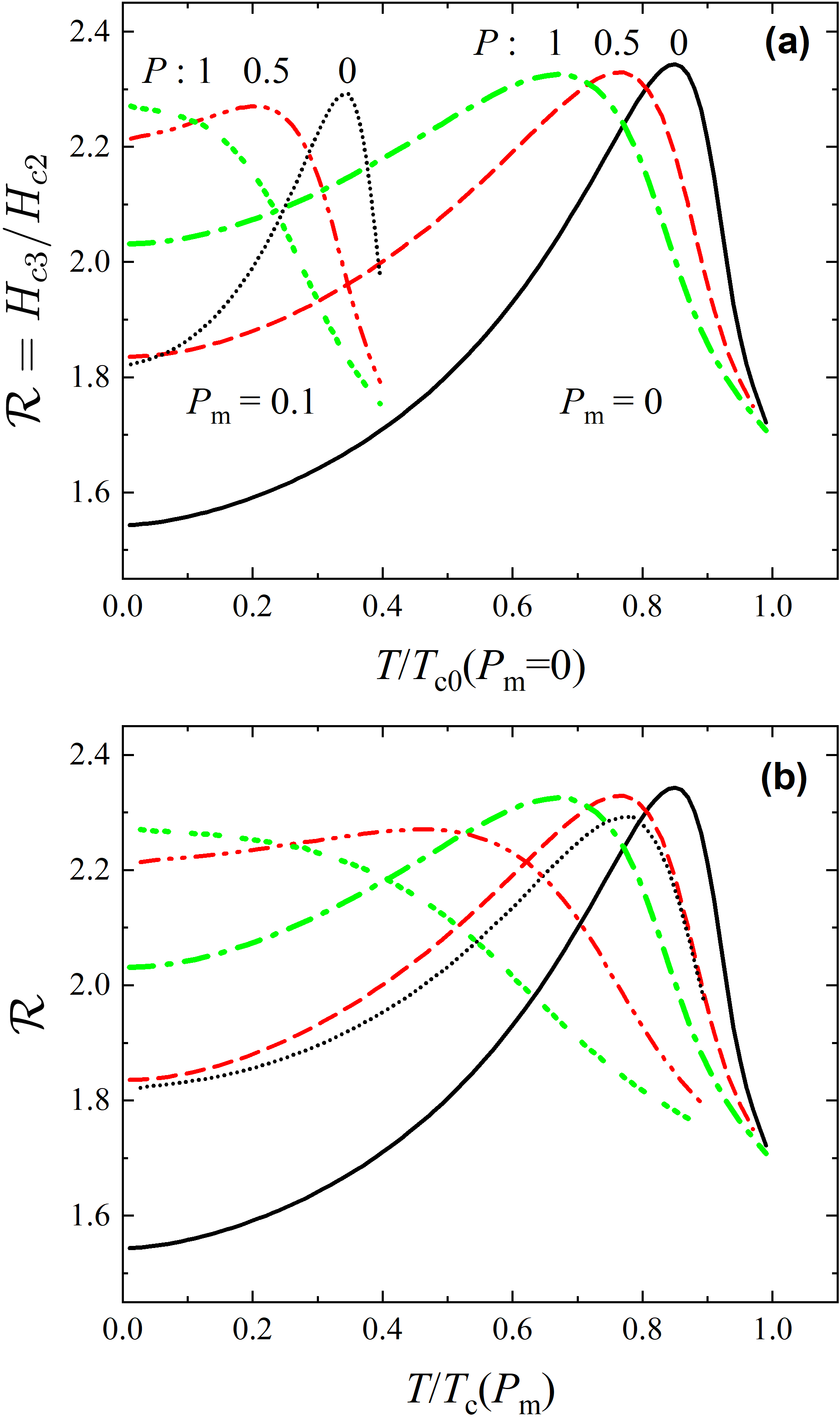}
\caption{(a)  ${\cal R}(P,P_m)=H_{c3}/H_{c2}$ vs. $T/T_{c0}$ for the scattering parameters indicated. (b) The same plotted vs. the actual reduced temperature $T/T_{c}(P_m)$.}
\label{f1}
\end{figure}


For  numerical work we recast the self-consistency equation  to dimensionless form:
\begin{eqnarray}
-\ln t=  \sum_{n=0}^\infty\left[ \frac{1}{n+1/2}-\frac{2tS}{2t(n+1/2)+P^+-SP^-}\right]. \nonumber\\ 
\label{s-c}
\end{eqnarray}


The calculated ratio ${\cal R}=H_{c3}/H_{c2}$ as function of the reduced temperature $T/T_{c0}$ for a few values of scattering parameters $P$ and $P_m$ is shown in the upper panel of Fig.\,\ref{f1}; the lower panel shows the same results plotted vs $T/T_c$. One can see that the maximum of  ${\cal R} $   shifts to lower $T$ with increasing transport scattering $P$. Effects of magnetic scattering are mostly in suppressing the actual critical temperature $T_c$ and larger values of ${\cal R}$ at low temperatures as compared to purely transport scattering.

\begin{figure}[tb]
\includegraphics[width=8cm,left] {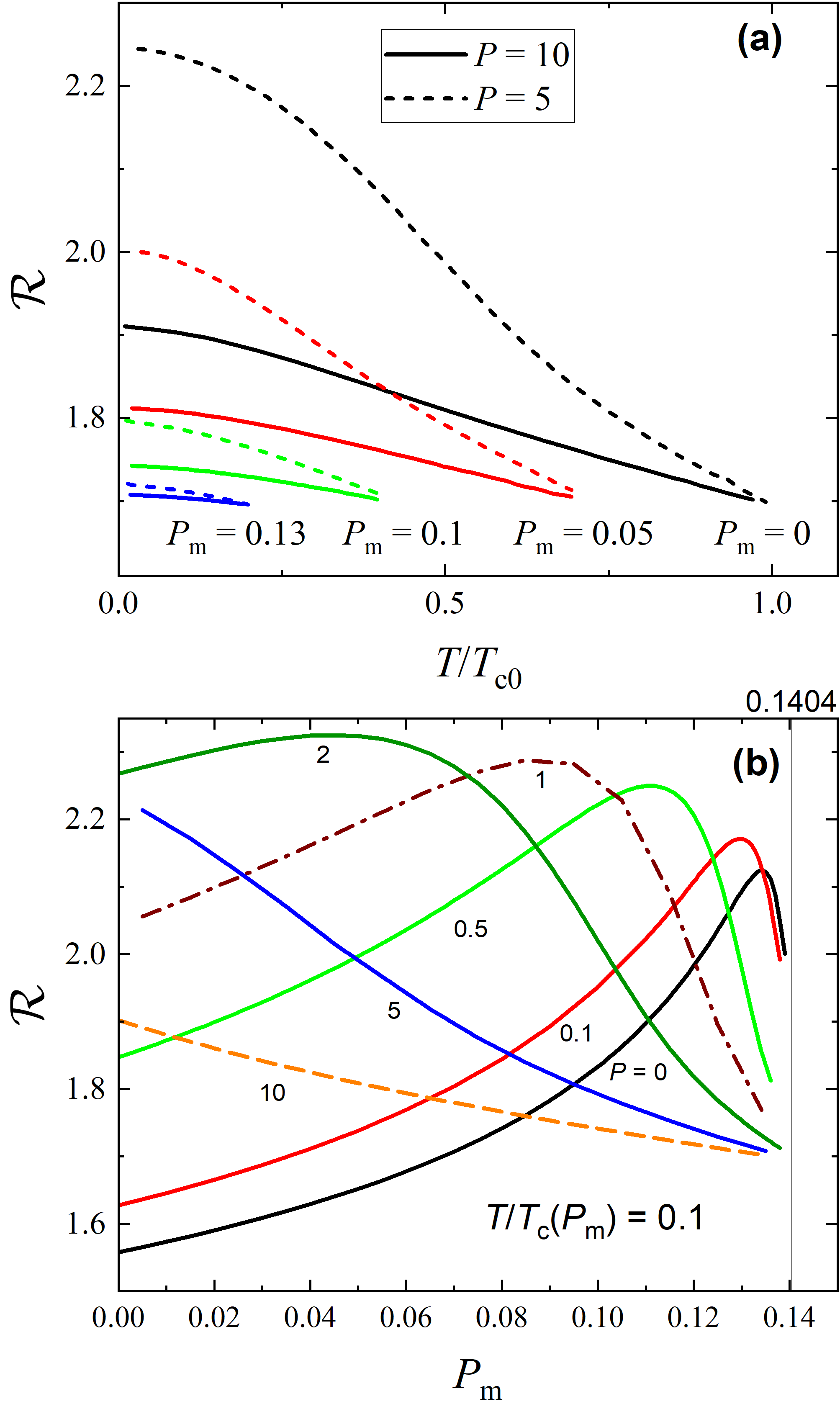}
\caption{ (a) ${\cal R}(T)=H_{c3}/H_{c2}$ vs. $T/T_{c0}$ for strong transport scattering and a set of $P_m$ values for magnetic scattering. (b) ${\cal R}$ vs. pair-breaking scattering parameter $P_m$ at a low reduced temperature, $T/T_c(P_m)=0.1$ for a set of $P$ values of transport scattering. Note: the maximum $P_m$ possible is 0.1404.}
\label{f2}
\end{figure}

It was shown in Ref.\,\cite{Levch} that in the absence of magnetic impurities and a strong transport scattering, the ratio ${\cal R}(T) $ flattens and the $T$ dependence disappears in the dirty limit in which ${\cal R}(T) \approx 1.7$ at all $T$s. This is due to the disappearing field dependence of the coherence length $\xi$ \cite{KN} in this limit. It is thus instructive to see that   the magnetic scattering does not change this qualitatively, see  the upper panel of Fig.\,\ref{f2}.

Hence, magnetic impurities do not change drastically the behavior of  $ H_{c3}$ relative to $ H_{c2}$. We find that within the isotropic s-wave theory, for the standard geometry of a half-space sample in a field parallel to the surface, the ratio ${\cal R}=H_{c3}/H_{c2}$ is within the window  $1.55\lesssim {\cal R}\lesssim 2.34$, regardless of temperature, magnetic or non-magnetic scattering.

    On the dirty side (with a strong transport scattering), the maximum of ${\cal R}(T)$ moves to $T\approx 0$ as is seen in the lower panel of Fig.\,\ref{f2}. Effects of pair breaking here are  not   drastic, even for the gapless situation ($0.128<P_m<0.14$, \cite{KPMishra}) we still have ${\cal R}\approx 1.7$. 

\section{Type of superconductivity and magnetic impurities}

 The basic Eqs.\,(\ref{E1}) and (\ref{selfcons0}) can be
obtained minimizing the energy functional as is done in the original Eilenberger's paper \cite{Eil}  for exclusively transport scattering:
\begin{eqnarray}
    \Omega&=&N(0)\left[\Delta^2\ln \frac{T}{T_{c0}}
 +2\pi T\sum_{\omega>0} \left(\frac{\Delta^2}{   \omega
}-\Big\langle I\Big\rangle\right)\right],\qquad\label{Omega}\\
I&=&2\Delta f +2 \omega (g-1)+ \frac{  f\langle f\rangle }{ 2\tau^-}+\frac{ g\langle g\rangle -1}{2 \tau^+}.\qquad
\label{I}
\end{eqnarray}
 The function $g$ in (\ref{I}) is an abbreviation for $\sqrt{1-f^2}$. 
 The free energy difference between superconducting and normal states is obtained by substituting solutions of Eqs.\,(\ref{E1})  in $\Omega$. In particular, taking account of the self-consistency equation, we obtain for the condensation energy density $F_c=F_n-F_s$:
\begin{eqnarray}
 \frac{F_c}{2\pi TN(0)}  
= \sum_{\omega>0} \Big\langle \Delta f +2 \omega (g-1)
+\frac{  f\langle f\rangle }{ 2\tau^-}+\frac{ g\langle g\rangle -1 }{2 \tau^+}\Big\rangle .\qquad 
 \label{Omega1}
\end{eqnarray}
 This expression reduces to the known BCS result for isotropic s-wave cases with or without magnetic impurities \cite{Maki}. For uniform zero-field state, the averaging  brackets can be omitted, and the scattering part is $-  f^2/\tau_m$. 
 Introducing dimensionless order parameter  $\delta=\Delta/2\pi T_{c0}$ one  has:
\begin{eqnarray}
\frac{F_c}{4\pi^2 T_{c0}^2N(0)}  = t \sum_{n=0}^\infty \left[ \delta f +t(2n+1 )  (g-1) -P_m f^2\right].\nonumber\\ 
 \label{nondim}
\end{eqnarray}
 The thermodynamic critical field follows
\begin{eqnarray}
H_c=\sqrt{8\pi F_c}=\left\{ 32\pi^3N(0)T_{c0}^2\right\}^{1/2} \,U(t)\,,
\end{eqnarray}
where 
\begin{eqnarray}
U(t)= 
 \left\{ t  \sum_{n=0}^\infty \Big[ \delta f +t(2n+1 )  (g-1) -P_m f^2\Big]\right\}^{1/2}. \qquad
\end{eqnarray}
  Hence, we have the dimensionless thermodynamic field:
\begin{eqnarray}
h_c=\frac{H_c \hbar^2v^2}{2\pi \phi_0T_{c0}^2} = \frac{ \hbar^2v^2}{ \phi_0T_{c0} }\sqrt{8\pi N(0)}\,U(t)\,. \end{eqnarray}

 The pre-factor by $U$, which is a characteristics of he clean material, can be expressed in terms of $GL$ parameter $\kappa_0$ for the {\it clean limit} \cite{Fetter}:
 \begin{eqnarray}
\kappa_0=\frac{3 \phi_0T_{c0} }{ \hbar^2v^2  \sqrt{ 7\zeta(3)\pi N(0)}}\,.
\end{eqnarray}
Thus, we obtain
\begin{eqnarray}
h_c(t,P_m)=\frac{3}{\kappa_0}\sqrt{\frac{8 }{7\zeta(3)}}\,U(t,P_m)\,.  
\label{hc}
\end{eqnarray}

To evaluate the condensation energy (\ref{nondim}) one first has to find $f$ and $\delta$ for given $t$ and $P_m$. For the uniform zero-field state these are solutions of the Eilenberger equation for $f$ and of the self-consistency equation. In our notation this system of two equations for $f(t,P_m)$ and $\delta(t,P_m)$ reads:
\begin{eqnarray}
\sqrt{1-f^2}(\delta-P_mf)-t(n+1/2)f=0\,,\label{eE1}\\
-\delta\ln t =  \sum_{n=0}^\infty \left(\frac{ \delta}{n+1/2} -t  f \right).  
 \label{system}
\end{eqnarray}
The system  can be solved numerically with the help of Wolfram Mathematica or MATLAB. Evaluation of $h_c(t,P_m)$ is then straightforward.

\begin{figure}[tb]
\includegraphics[width=8.7cm,left] {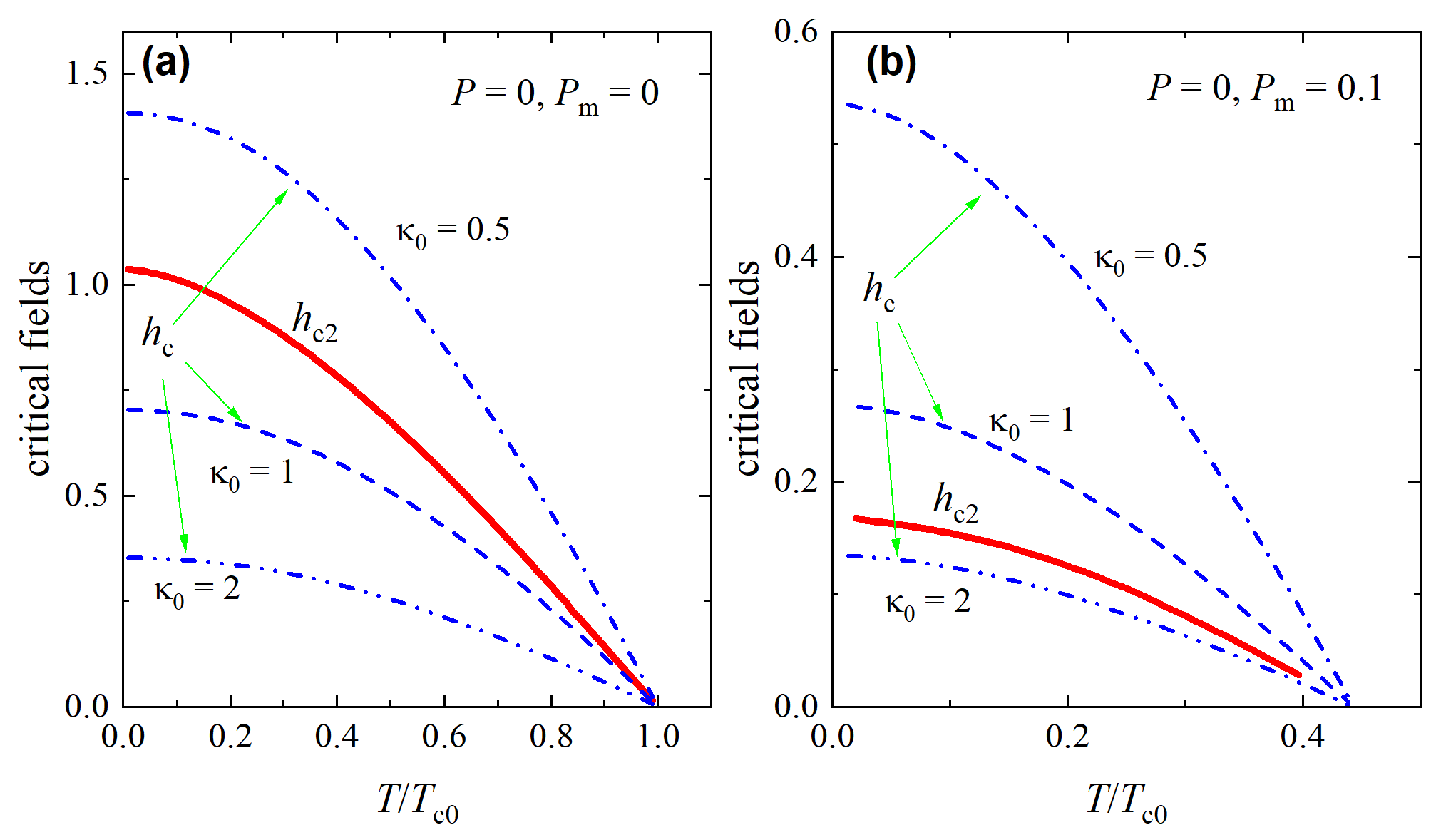}
\caption{$H_c$ (thin dashed lines) for the indicated values of Ginzburg-Landau parameter, $\kappa_0$, and $ H_{c2}$ (thick red line) in units $2\pi\phi_0 T_{c0}^2/\hbar^2v^2$ vs. $T/T_{c0}$. (a) Clean case, $P=P_m=0$.  (b) Strong magnetic scattering, $P_m=0.1$, in the absence of potential scattering, $P=0$. When a dashed blue line is above the solid red line, the material is a type-I superconductor.}
\label{f3}
\end{figure}

Results for $h_c$ and $h_{c2}$ are shown in the left panel of Fig.\,\ref{f3}
for the clean case. For the parameter $\kappa=0.5$, we have a type-I behavior, while the type-II is realized for $\kappa=1$ and 2, as should be since the boundary value is $\kappa_0=1/\sqrt{2}\approx 0.7$. Effect of the pair-breaking scattering is non-trivial; to see this we show the case of exclusively pair-breaking scattering, in which  for both $\kappa=0.5$ and $\kappa=1$ the material behaves as type-I since $h_c>h_{c2}$. 

One may say that the pair-breaking scattering pushes materials toward type-I, the conclusion we arrived at in \cite{I-II} in a different manner.
 
\section{Summary}

We have studied the effects of transport and pair-breaking scattering on the upper critical field $H_{c2}$, the thermodynamic critical field $H_{c}$, and the nucleation field $H_{c3}$ of surface superconductivity for the field   parallel to the plane surface of the half-space isotropic sample. We did not touch on questions of surface roughness, surface curvature, inhomogeneous distribution of impurities \cite{Sauls2019}, material anisotropy  \cite{Max}, etc. 

\begin{figure}[tb]
\includegraphics[width=8.7cm] {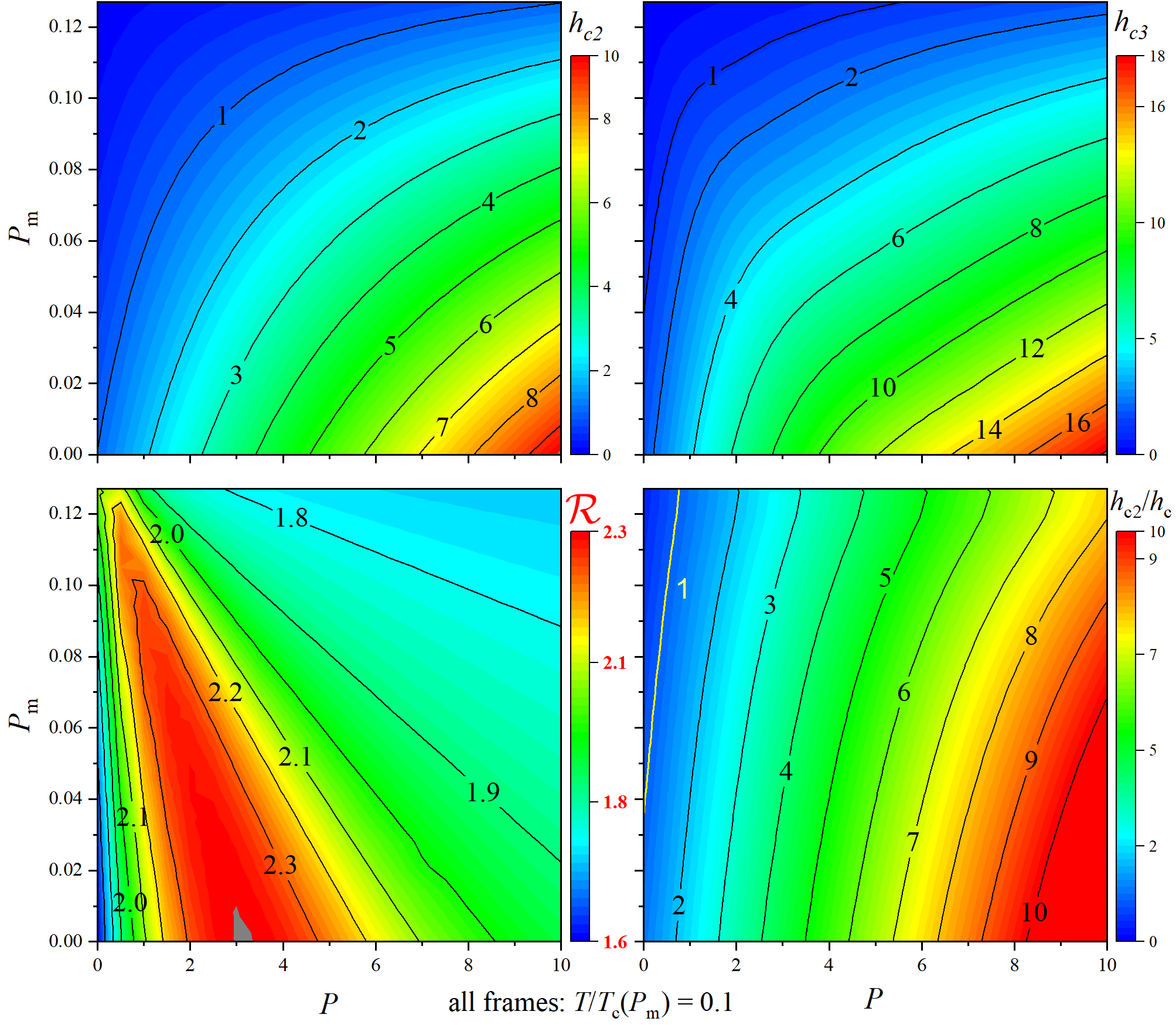}
\caption{ The summary of our results for $H_{c2}(P,P_m)$ (top left),  $H_{c3}(P,P_m)$ (top right), ${\cal R}(P,P_m)=H_{c3}(P,P_m)/H_{c2}(P,P_m)$ (lower left), and the ratio $H_{c2}(P,P_m)/ H_{c}(P,P_m)$  (lower right) at the same reduced temperature $T/T_c(P_m)=0.1$. All fields are in units of $2\pi \phi_0T_{c0}^2/\hbar^2v^2$.
} 
\label{f4}
\end{figure}

Whereas $H_{c2}$ is  suppressed by the pair-breaking scattering, $H_{c3}$ is found to be suppressed as well so that the ratio ${\cal R}=H_{c3}/H_{c2}$ is within the window  $1.55\lesssim {\cal R}\lesssim 2.34$  regardless of temperature, magnetic or non-magnetic scattering. We find that the magnetic impurities do not change qualitatively the behavior of the ratio ${\cal R} $ with changing temperature and  transport scattering: ${\cal R}(T) $ is  equal to SJDG value 1.695  at $T_c$ but increases on cooling, goes through a maximum at intermediate temperatures and then drops to a $P$ dependent value at $T=0$ \cite{Levch}.   The addition of magnetic impurities does not change this qualitative behavior, the suppression of the critical temperature notwithstanding. 

The thermodynamic critical field $H_c$, along with the condensation energy, is also suppressed by the pair-breaking scattering, but depending on material parameters and temperature the speed of this suppression could be larger or smaller than that of $H_{c2}$. On the other hand, the value of  $H_{c}$ relative to $H_{c2}$ is crucial for the type of emerging superconductivity: type-I for  $H_{c}>H_{c2}$, whereas type-II for  $H_{c}<H_{c2}$. A possibility of changing the type of material superconductivity by changing the concentration of magnetic impurities has also been discussed in \cite{I-II}. 

The summary of our results for the critical fields at a low temperature is given in Fig.\,\ref{f4}. 

Using our routine of calculating the condensation energy, we looked at the phase transition between gapped and gapless superconductivity, the question   recently attracted the community's attention. We found that the third derivative of the free energy with respect to the pair-breaking parameter has a singular discontinuous jump at $T=0$ predicted in \cite{Varlamov, Var1}.  The transition,  however, broadens at finite $T$,  see Appendix B.

We  mention yet another example of the 2nd order phase transition that can be treated within the same formal scheme as $H_{c3}$. This is the problem of nucleation of superconductivity  in thin   films  in parallel applied field \cite{KN}. The boundary condition $\nabla_n\Delta(\bm r)=0$ should now be obeyed at both film surfaces and the emerging  state, being physically similar to that of surface superconductivity of SJDG, can even be nucleated at $T>T_c$ at a non-zero magnetic field. The $T_c(H)$ enhancement  in this geometry had been observed \cite{Schlom,Xie}, but a careful investigation of this intriguing possibility is still to be done. 
 
 \section{Acknowledgments}
 
The authors are grateful to Jim Sauls, Alex Levchenko and Alex Gurevich for helpful discussions. This work was supported by the U.S. Department of Energy (DOE), Office of Science, Basic Energy Sciences, Materials Science and Engineering Division. Ames Laboratory  is operated for the U.S. DOE by Iowa State University under contract \# DE-AC02-07CH11358. 
RP acknowledges support by the DOE National Quantum Information Science Research Centers, Superconducting Quantum Materials and Systems Center (SQMS) under contract No. DE-AC02-07CH11359.

\appendix

\section{The sum $\bm S$ in the presence of magnetic impurities}

The solution $f$ of Eq.\,(\ref{E1}) can be written as
\begin{eqnarray}
 f &=&  (2\omega^+ +\bm v\bm\Pi)^{-1} (F/\tau^- + 2\Delta) \nonumber\\
 &=&   \int_0^\infty d\rho e^{-\rho(2\omega^+ +\bm v\bm\Pi)} (F/\tau^- + 2\Delta) \,. 
\label{f}
\end{eqnarray}
Taking the Fermi surface average we get
\begin{eqnarray}
F=\frac{1}{\tau^-}   \int_0^\infty d\rho e^{-2 \omega^+\rho}\left\langle e^{-\rho\bm v\bm\Pi}\right\rangle(F+ 2\Delta\tau^- ) \,. \qquad
\label{Fa}
\end{eqnarray}
The term $\langle...\rangle$ does not contain the scattering parameters, hence it is the same as that calculated in \cite{85} for the clean case:
\begin{eqnarray}
\left\langle e^{-\rho\bm v\bm\Pi}\tilde F\right\rangle = \sum_{m,j}\frac{(-q^2)^j}{(m!)^2j!}\frac{(2\mu)!!}{(2\mu+1)!!} \left(\frac{\rho v}{2}\right)^{2\mu}(\Pi^+)^m(\Pi^-)^m\tilde F.  \nonumber\\
\label{average}
\end{eqnarray}
Here $\tilde F=F+2\Delta\tau^-$, $\mu=m+j$, and $\Pi^{\pm}=\Pi_x\pm i \Pi_y$.
After integrating over $\rho$, one obtains from Eq.\,(\ref{Fa})
\begin{eqnarray}
&&F=\frac{1}{2\omega^+\tau^-}
  \sum_{m,j}\frac{(-q^2)^j}{j!(2\mu+1)}\left(\frac{ \mu !}{m!}\right)^2 \left(\frac{\ell^+}{\beta^+}\right)^{2\mu} (\Pi^+)^m(\Pi^-)^m\tilde F \nonumber\\
&&\ell^+=v\tau^+,\quad  \beta^+=1+2\omega \tau^+\,.
\label{average1}
\end{eqnarray}
 One can check that if no magnetic impurities are involved, this reduces to Eq.\,(12) of \cite{85}. Using commutation properties of operators $\Pi^\pm$ in uniform field, one manipulates 
\begin{eqnarray}
 (\Pi^+)^m(\Pi^-)^m\tilde F = \tilde F \prod_{i=1}^{m} [k^2+(2i-1)q^2]
\label{algebra}
\end{eqnarray}
and obtains:
\begin{eqnarray}
F&=&\Delta \frac {2\tau^- S}{2\omega^+\tau^- -S}    \,,\nonumber\\
S&=&  \sum_{m,j}\frac{(-q^2)^j}{j!(2\mu+1)}\left(\frac{ \mu !}{m!}\right)^2 \left(\frac{\ell^+}{\beta^+}\right)^{2\mu}   \prod_{i=1}^{m} [k^2+(2i-1)q^2] \nonumber\\ 
\label{FS}
\end{eqnarray}

\begin{figure}[h]
\includegraphics[width=8.3cm] {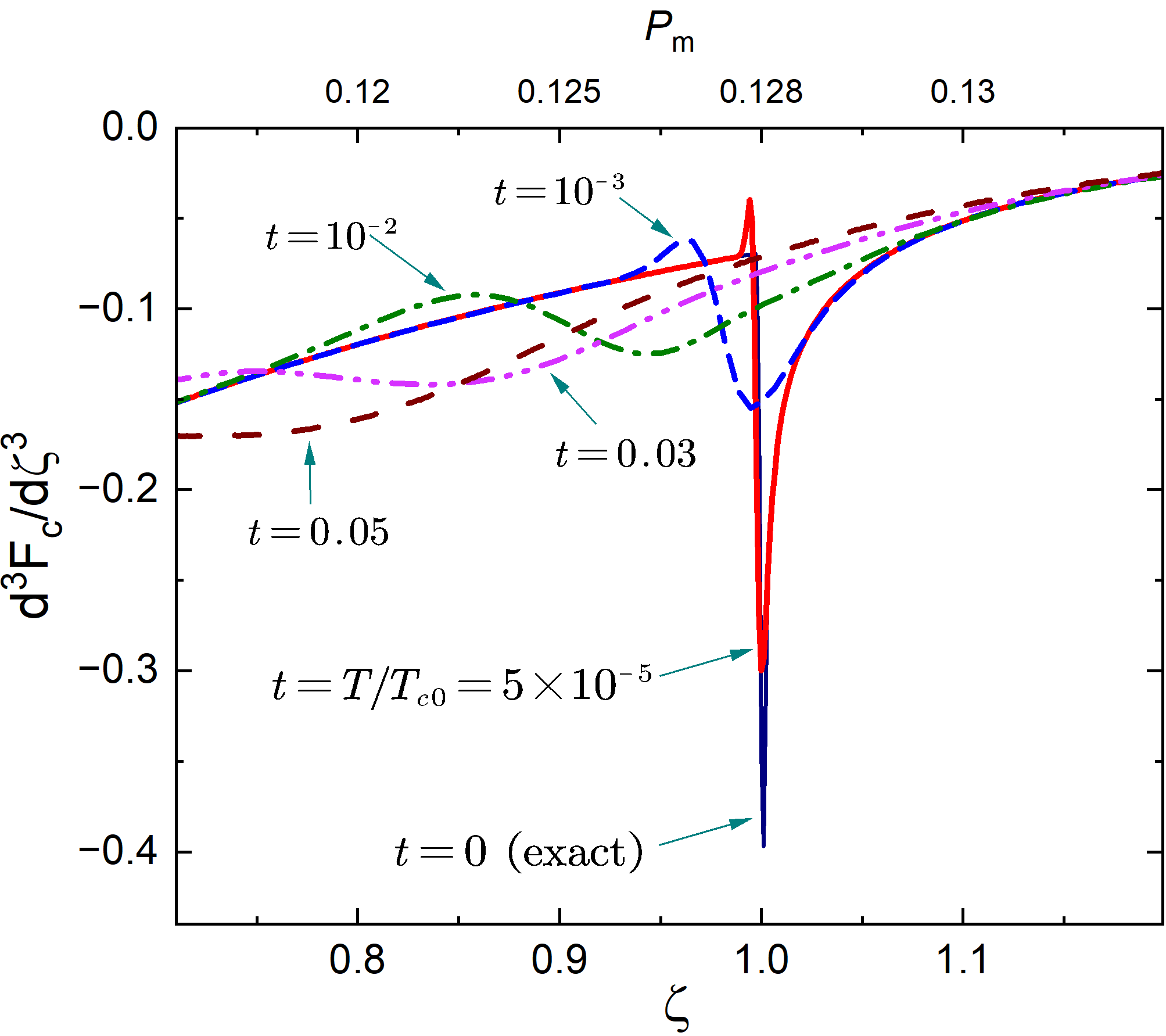}
\caption{The third derivative of the free energy at a set of low temperatures $t=T/T_{c0}=10^{-5}, 10^{-3}, \,{\rm and} \,\,10^{-5}$ vs the pair-breaking parameter $ \zeta = \hbar/\Delta\tau_m=(2\pi T_{c0}/\Delta)P_m$. In addition, the curve at $T=0$ obtained using the exact Eq.(71) from Maki \cite{Maki}.  $ \zeta = 1$ and $P_m=0.128$ correspond  to the transition to gapless state at $T=0$. In this calculation the sum over $n$ was extended up to $N_{max}=10^5$.
} 
\label{f5}
\end{figure}

\section{Condensation energy vs pair-breaking parameter $\bm {P_m}$}

Recently   the character of  the quantum phase transition between gapped and gapless superconductivity at $T=0$ \cite{Var1}    when the pair-breaking scattering parameter $P_m$ varies through the   value $e^{-\pi/4-\gamma}/2=0.128$ (see e.g. \cite{KPMishra}). 
Remarkably, It turned out  that the superconducting density of states as function of energy $\omega$ and of the order parameter $\Delta(P_m)$ undergoes a topological transition at  $P_{m}=0.128$. In Ehrenfest classification, the transition can be considered as of the 2.5-order, at which the third derivative of the free energy $\partial^3 F_c/\partial P_m^3$ is singular.

Although this question is out of main subject of this paper, we utilize here  the   functional (\ref{Omega}), (\ref{I}), and the condensation energy $F_c(T,P_m)$ of Eq.\,(\ref{Omega1}) derived above and valid for any $T$ and any scattering parameters $P,P_m$.  

One should mention that solving a coupled system of Eqs.\,(\ref{eE1}   and (\ref{system} is not a trivial task. The lower the temperature, the more Matsubara summations are required. Initially, calculations were conducted with the help of Wolfram Mathematica, however, it could not handle the lowest temperatures. Final calculations were performed within MATLAB that still required at least 100,000 summations to obtain the reported  results (interested readers are welcome to contact the authors for further technical details).

As can be seen in Fig.\,\ref{f5}, temperature affects the behavior of the third derivative of the condensation energy  dramatically when compared to the exact result at $T=0$ obtained using the Eq.(71) from Maki \cite{Maki}.
Our calculations confirm the existence of a very sharp discontinuity of  $\partial^3F_c(T,P_m)/\partial P_m^3$ at $T\to 0$ at $P_{m}=0.128$ (or at $ \zeta = \hbar/\Delta\tau_m=(2\pi T_{c0}/\Delta)P_m=1$ in notations of Ref.\,\cite{Varlamov}). However, we were unable to confirm the claim  that the discontinuity is preserved at non-zero temperatures  \cite{Varlamov}; our calculation shows that the singularity in $F'''_c(P_m)$ broadens with increasing $T$. In fact, at finite temperatures, the singularity disappears while its trace moves to lower scattering rates $P_m$. As expected, this confirms  that the critical magnetic scattering rate for a transition to the gapless state decreases starting from   $P_m=0.128$ at $T=0$ to lower values at higher temperatures \cite{Amb}.


\begin{thebibliography}{99}

\bibitem{Gurevich2012}A. Gurevich, Reviews of Accelerator Science and Technology \textbf{05}, 119 (2012).

\bibitem{Liarte2017}D. B. Liarte, S. Posen, M. K. Transtrum, G. Catelani, M. Liepe, J. P. Sethna, Superconductor Science and Technology \textbf{30}, 33002 (2017).

\bibitem{Devoret2013}M. H. Devoret, R. J. Schoelkopf, Science \textbf{339}, 1169 (2013).

\bibitem{Rigetti2012}C. Rigetti, J. M. Gambetta, S. Poletto, B. L. T. Plourde, J. M. Chow, A. D. C\'{o}rcoles, J. A. Smolin, S. T. Merkel, J. R. Rozen, G. A. Keefe, M. B. Rothwell, M. B. Ketchen, M. Steffen,  \prb \textbf{86}, 100506 (2012).

\bibitem{Sauls2019}V. Ngampruetikorn and J. A. Sauls, Phys. Rev. Res. {\bf 1}, 12015 (2019).

\bibitem{DG} D. Saint-James and P.G. De Gennes, Phys. Lett. {\bf 7}, 306 (1963).  

\bibitem{SJ}D. Saint-James, G. Sarma, E. J. Thomas \textit{Type II
Superconductivity}, Pergamon, Oxford, New York, 1969.

\bibitem{DGbook}P.-G. de Gennes, {\t Superconductivity of Metals and
Alloys} (Advanced Book Program, Perseus Books, Reading, Mass., 1999). 

\bibitem{Tomasch} W. J. Tomasch and A. S. Joseph, \prl {\bf 12}, 148 (1964).

\bibitem{hemp} C. F. Hempstead and Y. B. Kim, \prl {\bf 12}, 145 (1964).

\bibitem{Strong} M. Strongin, A. Paskin, O. F. Kammerer, and M. Garber, \prl {\bf 14}, 362 (1965).

\bibitem{Max}V. G. Kogan, J. R. Clem, J. M. Deang, and M. D. Gunzburger,
\prb {\bf 65}, 094514 (2002).

 \bibitem{Levch}  Hong-Yi Xie, V. G. Kogan,  M. Khodas,  and A. Levchenko,    \prb   {\bf 96}, 104516 (2017).  

 \bibitem{KP-2013}V. G. Kogan and R. Prozorov, \prb {\bf 90}, 180502(R) (2014).


\bibitem{Eil}G. Eilenberger, Z. Phys. {\bf  214}, 195 (1968).

\bibitem{KPMishra}V. G. Kogan, R. Prozorov, and V. Mishra, \prb {\bf 88}, 224508 (2013).

\bibitem{HW}E. Helfand  and N. R. Werthamer, Phys. Rev. {\bf 147}, 288 (1966). 

\bibitem{85}V. G. Kogan, \prb {\bf 32}, 139 (1985).

\bibitem{KN}V. G. Kogan and N. Nakagawa, \prb {\bf 35}, 1700 (1987).

\bibitem{Maki} K. Maki, in {\it Superconductivity}, edited by R. D. Parks, Vol. 2
(Marcel Dekker, New York, 1969).

\bibitem{Fetter} A. Fetter and P. C. Hohenberg, in {\it Superconductivity}, edited by R. D. Parks, Vol. 2
(Marcel Dekker, New York, 1969).

 \bibitem{I-II}V. G. Kogan and R. Prozorov, \prb {\bf 88}, 024503 (2013).
 
\bibitem{Varlamov}Y. Yerin, A. A. Varlamov, and C. Petrillo, arXiv:2205.13951 (2022).

\bibitem{Var1} Y. Yerin, C. Petrillo, A. A. Varlamov, Sci-
Post Core, {\bf 5}, 009 (2022) .

\bibitem{Schlom}H. J. Gardner, A. Kumar, Liuqi Yu, Peng Xiong, M. P. Warusawithana, 
Luyang Wang, O. Vafek, and D. G. Schlom, Nature Physics, {\bf 7}, 895 (2011).

\bibitem{Xie}Yonghui Ma, Jie Pan, Chenguang Guo, Xuan Zhang, Lingling Wang, Tao Hu, Gang
Mu, Fuqiang Huang, and Xiaoming Xie, arXiv:1712.07763 (2017).

\bibitem{Amb}V. Ambegaokar and A. Griffin, Phys. Rev. {\bf 137}, A1151 (1965).

\end{thebibliography}
\end{document}